\documentstyle[aps,epsf,eqsecnum]{revtex}
\begin{document}
\preprint{UAHEP 98?}
\draft
\title{Energy Absorption by the Dilaton Field around a Rotating
Black Hole in a Binary System}
\author{R. Casadio}
\address{Dipartimento di Fisica, Universit\`a di Bologna, and \\
I.N.F.N., Sezione di Bologna, \\
via Irnerio 46, 40126 Bologna, Italy}
\author{B. Harms}
\address{Department of Physics and Astronomy,
The University of Alabama\\
Box 870324, Tuscaloosa, AL 35487-0324}
\maketitle
\begin{abstract}
In this paper we analyze a binary system consisting of a star and a
rotating black hole.
The electromagnetic radiation emitted by the star interacts with
the background of the black hole and stimulates the production
of dilaton waves.
We then estimate the energy transferred from the electromagnetic
radiation of the star to the dilaton field as a function of
the frequency.
The resulting picture is a testable signature of the existence
of dilaton fields as predicted by string theory in an astrophysical
context.
\end{abstract}
\section{Introduction}
\label{intro}
The theory of strings and membranes is a very attractive candidate
for the quantum theory of gravity, however, few results have been
derived which can be tested at present.
In this approach, our Universe is supposed to be described by the
low energy effective action of string theory.
The latter can be written in terms of the fields associated with
the massless excitations of extended objects, among which there is
a scalar field (the dilaton) (see, {\em e.g.},
Ref.~\cite{polchinski}).
In a series of papers \cite{hl1,hl2,chl1,hl3,hl4,hl5,hl6,hl7} we
have explored the viewpoint that quantum black holes are massive
excitations of extended objects and hence are elementary particles.
Our goal is thus to obtain expressions which can be compared to
experimentally measurable quantities in order to test the idea that
quantum black holes are extended objects.
In all likelihood the scalar component of gravity, provided it
does not acquire a mass from quantum corrections \cite{polchinski},
will have a coupling to electromagnetic waves which is as weak as
that of the tensor component.
Therefore the best place to look for the effect of the long range
dilaton on electromagnetic waves is in the neighborhood of a black
hole, which we consider to be composed of quantum black holes.
\par
In Ref.\cite{knd}, starting from the low energy effective action
describing the Einstein-Maxwell theory interacting with a dilaton
$\phi$ in four dimensions $(G = 1)$ \cite{frame},
\begin{eqnarray}
S = {1\over16\,\pi}\,\int d^4x\,\sqrt{-g}\,\left[R-{1\over2}\,
(\nabla\phi)^2-e^{-a\,\phi}\,F^2\right]
\ ,
\label{action}
\end{eqnarray}
we obtained the static solutions of the field equations for a
Kerr-Newman dilaton (KND) black hole rotating with arbitrary angular
momentum by expanding the fields in terms of the charge-to-mass ratio of
the black hole.
In \cite{kndw,kndw2} we used these solutions, which appear as a
background in the linear wave equations, to investigate the effects
of the background dilaton on the propagation of various spin waves in
the vicinity of a rotating charged black hole.
In particular, we showed that the different wave modes disentangle
at a given order in the charge-to-mass expansion and this allowed us
to study the electromagnetic waves analytically up to second order
in the charge-to-mass ratio.
\par
The point of view we take in the present paper is more
phenomenological in that we focus on the search for observable
effects.
We want to model a binary system which is composed of a rotating
black hole and a star and estimate the perturbation that would be
induced on the electromagnetic spectrum of the star, as detected by a
distant observer, by the existence of a scalar component of gravity.
This does not require the presence of a static dilaton field, nor does the
black hole itself need to be electrically charged.
All that is needed is a background electromagnetic field whose source
could be, {\em e.g.}, in the accreting disk which we consider as part
of the black hole component of the binary system.
\par
In particular, we shall be interested in the case when the plane
containing the system is roughly parallel to the direction of
observation since then we expect stronger contributions.
This configuration implies that when the star is in front of the
black hole, the background of the black hole negligibly
affects the radiation coming from the star.
On the other hand, when the star is going behind the black hole,
the radiation emitted by the star will travel across a region where
the static electromagnetic field is stronger.
The leading order effect is then the interaction between electromagnetic
waves and the electromagnetic background which produces dilaton waves,
thus carrying away a certain amount of energy.
This energy can be transferred from dilaton modes back in the form
of electromagnetic radiation or to gravitational waves, both cases being
next order in the dilaton coupling constant and in the charge-to-mass
ratio in our approximation scheme \cite{kndw,kndw2}.
Therefore we shall assume that the dilaton waves retain most of their
energy and simply result in a permanent loss from observation.
We shall then study the dilaton wave equation and estimate the energy
lost by the radiation of the star so that, by comparing the spectra
corresponding to the two different relative positions of the star and
the black hole, one can infer (or disprove) the existence of scalar
gravitational excitations.
\par
We begin in Section~\ref{system} with a detailed description of the system
under study, the governing equations and the approximations we will assume
in order to manage the equations analytically.
In Section~\ref{waves} we obtain expressions for the dilaton waves generated
by the radiation emitted by the star.
Finally, in Section~\ref{flux} we estimate the energy flux at large distance
and in Section~\ref{close} discuss our results.
For the notation and the complete description of the background of the
rotating black hole we refer to \cite{knd}; for the derivation of linear
perturbations the main reference is \cite{kndw}.
\section{The binary system}
\label{system}
We consider a binary system made of a star and a rotating black hole.
If the black hole is electrically charged, then from string theory
one expects a non-trivial background dilaton field \cite{knd}.
However, as we mentioned in the Introduction, the charge does not need
to be located in the black hole singularity itself.
In fact, the metric we shall be using to describe the black hole (as
well as any other black hole metric) works well for the exterior of any
rotating charged distribution of matter.
Given any charge distribution in the matter just outside the horizon,
one could then approximate the true metric everywhere by pasting together
regions where the metric is given by Eq.~(\ref{metric}) below with
different values for its parameters.
For the sake of simplicity, in the following we shall consider only
one set of parameters for all points in the region of interest.
\par
We shall also assume that the presence of the star does not affect the
geometry of space-time significantly in the region near the outer
horizon of the black hole (denoted by $r_+$) where the static
electromagnetic field is presumably stronger.
This means that either the mass of the star is much smaller than the
black hole ADM mass $M$ or that the star is distant enough from the
black hole.
In both cases we approximate the background metric in the region of
interest by \cite{knd}
\begin{eqnarray}
ds^2 = - \Psi\,(dt - \omega\,d\varphi)^2
+{\Delta\,\sin^2\theta\over\Psi}\,(d\varphi)^2
+\rho^2\,\left[{(dr)^2\over\Delta}+(d\theta)^2\right]
\ ,
\label{metric}
\end{eqnarray}
in which $x^0=t$, $x^1=r$, $x^2=\theta$, $x^3=\varphi$ are Boyer-Lindquist
coordinates centered on the black hole.
The explicit expressions of the functions $\Psi=\Psi(r,\theta)$,
$\omega=\omega(r,\theta)$ and $\Delta=\Delta(r)$ in terms of
the mass $M$, charge $Q$ and angular momentum $J=\alpha\,M$
of the hole coincide (at order $Q^2/M^2$) with the Kerr-Newman (KN)
metric
\cite{kndw,chandra}
\begin{eqnarray}
\Delta&=&r^2-2\,M\,r+\alpha^2+Q^2
\nonumber \\
\rho^2&=&r^2+\alpha^2\,\cos^2\theta
\nonumber \\
\Psi&=&-(\Delta-\alpha^2\,\sin^2\theta)/\rho^2
\nonumber \\
\omega&=&-\alpha\,\sin^2\theta\,[1+\Psi^{-1}]
\ .
\label{KN-metric}
\end{eqnarray}
\par
Another basic observation is that, although the system is approximately
axisymmetric, if the star revolves along a circular orbit around the
black hole, the pattern of the radiation emitted by the star does not
share this symmetry, which renders the analysis extremely involved.
In fact, following standard Refs.~\cite{chandra,teukolsky}
the perturbations on the background (\ref{metric}) have
been studied by performing a decomposition into normal modes with
angular and radial parts according to this symmetry
\cite{kndw,kndw2}.
Hence, a flux of radiation emitted by the star, which is approximately
spherical with respect to the star, should first be decomposed into
normal modes on the black hole background, then propagated in the
region near the black hole and finally reconstructed at the observation
point (far away from the black hole).
\par
\begin{figure}
\centerline{\epsfysize=240pt\epsfbox{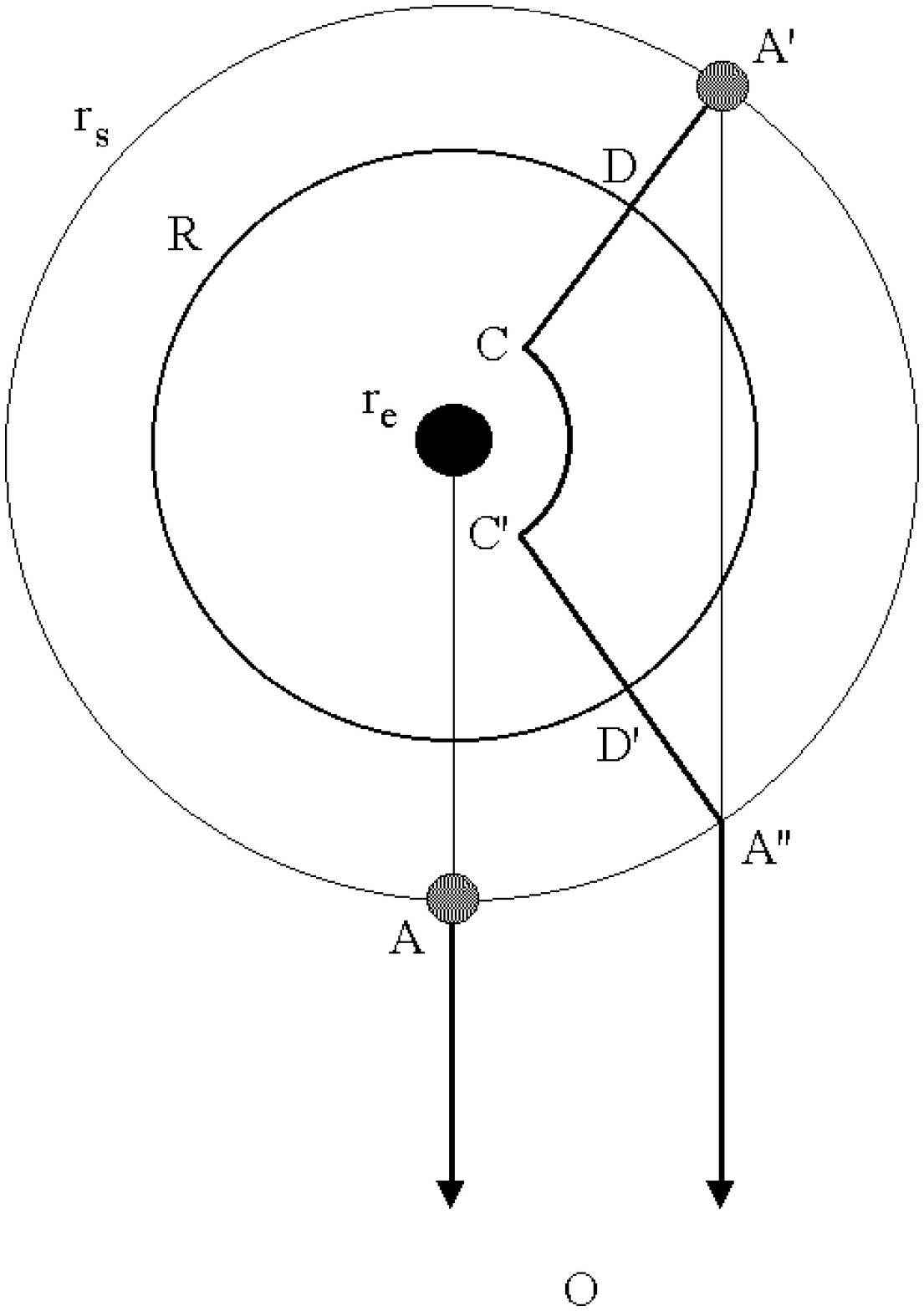}}
\caption{The binary system with the light paths as described in the
text.}
\label{sys}
\end{figure}

In order to speed up the analysis and estimate the order of magnitude
of the leading effects, we then focus on a particular configuration
of the binary system (Fig.~\ref{sys}).
As mentioned in the Introduction, we are interested in orbits of the
star which lie roughly along the line of sight of the observer in order
to produce occultations.
The simplest case is thus to take the orbit of the star, $r_s=r_s(t)$,
in the equatorial plane $\theta=\pi/2$ and assume that it also contains
the point of observation $O$.
The star must also be sufficiently away from the {\em stationary limit},
that is $r_s>r_e\equiv M+\sqrt{M^2-Q^2-\alpha^2\,\cos^2\theta}>
r_+\equiv M+\sqrt{M^2-Q^2-\alpha^2}$ (for $\theta=\pi/2$),
so that it does not affect the background geometry appreciably.
Further, the region where the interaction among the various fields
is relevant is assumed for convenience to be bounded by the radius
$R\sim r_e$.
Indeed, the parameter $R$ is not related to any basic physical process,
rather it is introduced for the purpose of taking into account the finite
precision of the measuring devices.
The use of such a fictitious parameter follows from the fact that,
because of the dependence of the background fields on the distance
from the hole, the intensity of the waves produced by the interactions
among the various fields falls off as a positive power of $1/r$, where
$r$ is the position at which the interaction takes place
\cite{kndw,kndw2}.
As such, $R$ will be kept unspecified and can be taken to infinity
or determined at the end by comparing the magnitude of the computed
effects with the precision of the measuring apparatus.
This will be further discussed in section~\ref{flux}.
\par
Regardless of the actual shape of the orbit of the star we shall consider
the following two cases:
\par\noindent
{\em i)} the star is at a point $A$ between the observer $O$ and the black
hole;
\par\noindent
{\em ii)} the star is at $A'$, being occultated by the black hole.
\par
In {\em i)} the relevant electromagnetic radiation emitted by the star
travels along the line $AO$ and will be approximated by outgoing modes
$\phi_i^{out}(r)$ ($i=1,2,3$) of the Maxwell equations to leading order
in the large $r$ expansion (see next section for the notation and
definitions).
This is a good approximation since the line $AO$ also contains the black
hole, so that deviations from axial symmetry are soft, and $r\ge r_s$ is
large.
Also, since $r$ is presumably bigger than $R$ we assume the electromagnetic
radiation propagates freely to the observer (on a Kerr background
\cite{kndw,chandra,teukolsky}).
\par
In {\em ii)} the path $A'O$ breaks the axial symmetry.
We then approximate $A'O$ with a new path $A'CC'A''O$ and, at first,
neglect the contribution from the arc $CC'$, so that the electromagnetic
radiation always travels along radial lines with respect to the black
hole and the normal modes studied in \cite{kndw} can be used
(more accurate estimates of the loss occurring along $CC'$ can be
obtained from the analysis carried on in section~\ref{nh}).
In particular, along $A'C$ we consider free ingoing modes $\phi_i^{in}$
and along $C'O$ free outgoing modes $\phi_i^{out}$ on the Kerr background.
Since the portion of the trajectory between $D$ and $D'$
lies inside $R$, this is where the electromagnetic radiation
generates other kinds of waves, thus dissipating energy.
\par
A fundamental difficulty in dealing with waves in the axisymmetric
space-time of a charged black hole is that electromagnetic, dilaton
and gravitational linear waves do not decouple.
However, as we mentioned in the Introduction, we have shown in
Refs.~\cite{kndw,kndw2} that it is indeed possible to disentangle the
various waves by expanding in the ratio $Q/M$.
In this way one finds that each linear wave mode of a given kind at a
given order satisfies an equation which contains only the background and
wave modes determined at lower orders.
This allows one to compute recursively every order in the $Q/M$ expansion
(at least in principle, since we know the background only up to order
$Q^3/M^3$) and this will be implemented in the following section for
the case at hand.
\par
As a final remark, we warn the reader that, since we consider just
one light path at a time, the gravitational lensing of the light coming
from the star is not accounted for.
The latter is an important relativistic effect which is expected to lead
to an increase of the luminosity of the occulting star \cite{occ}.
Therefore, a detailed balance between the loss of energy we shall
compute and the gravitational focusing requires the same
analysis  to be repeated for all possible light paths emanating from the star and
ending in $O$ and their contributions to be summed.
Of course, this program is very difficult to carry out analytically and we leave it
for future developments.
\section{Linear perturbations of the KND solution}
\label{waves}
Our input will be given by the electromagnetic waves coming from the star
and by examining all wave equations in Ref.~\cite{kndw} one easily
realizes that the leading effect in the $Q/M$ expansion is the interaction of
ingoing (along $DC$) and outgoing (along $C'D'$) electromagnetic waves
$\phi_i^{(1,0)}$ with the (static) electromagnetic background which
produces dilatonic waves $\phi_{in/out}^{(1,1)}(r)$.
Then, the first step is to compute the solutions of Maxwell's wave
equations at lowest order in $Q/M$ and use these solutions as sources for the
dilaton waves.
Since we are aiming at computing ratios between the amplitude of the
radiation emitted by the star and the amplitude of the dilaton waves
generated by the former, we shall not need to take into account the
proper normalization of the solutions.
\par
In order to make the present paper self-consistent we now proceed
to summarize the notation and review the relevant wave
equations for the electromagnetic and dilaton field,
\begin{eqnarray}
\left\{\begin{array}{l}
\nabla^i(e^{-a\,\phi}\,F_{ij}) = 0  \\
\nabla_{[ k}F_{ij]} = 0 \; , \\
\end{array}\right.
 \ \ {\rm (Maxwell)}
 \nonumber
\\
 \nonumber \\
\nabla^2\phi = -a\,e^{-a\,\phi}\,F^2 \ \ \ \ {\rm (dilaton)}
\ ,
\label{feq}
\end{eqnarray}
where $\nabla$ is the covariant derivative with respect to the metric
(\ref{metric}).
\par
We will use the standard Newman-Penrose null tetrad vectors \cite{NP}
for the KN metric \cite{chandra}
\begin{eqnarray}
l^i&=&{1\over\Delta}\,\left(r^2+\alpha^2,+\Delta,0,\alpha\right)
\nonumber \\
n^i&=&{1\over2\,\rho^2}\,\left(r^2+\alpha^2,-\Delta,0,\alpha\right)
\nonumber \\
m^i&=&{1\over\sqrt{2}\,\bar\rho}\,
\left(i\,\alpha\,\sin\theta,0,1,i\,\csc\theta\right)
\ ,
\label{tetrad}
\end{eqnarray}
where $\bar\rho\equiv r+i\,\alpha\,\cos\theta$ and
$\bar\rho^*\equiv r-i\,\alpha\,\cos\theta$.
Then the spin coefficients are represented by the following Greek letters
\begin{eqnarray}
&\kappa=\sigma=\lambda=\nu=\epsilon=0&
\nonumber \\
&\tilde\rho=-{1\over\bar\rho^*}\ ,\ \ \
\beta={\cot\theta\over 2\,\sqrt{2}\,\bar\rho}\ ,\ \ \
\pi={i\,\alpha\,\sin\theta\over\sqrt{2}\,(\bar\rho^*)^2}\ ,\ \ \
\tau=-{i\,\alpha\,\sin\theta\over\sqrt{2}\,\rho^2}&
\nonumber \\
&\mu=-{\Delta\over2\,\rho^2\,\bar\rho^*}\ ,\ \ \
\gamma=\mu+{r-M\over2\,\rho^2}\ ,\ \ \
\tilde\alpha=\pi-\beta^*
\ .&
\label{KN-spin}
\end{eqnarray}
The Maxwell field can be described by three
complex quantities,
\begin{eqnarray}
\phi_0&=&F_{ij}\,l^i\,m^j
\nonumber \\
\phi_1&=&{1\over 2}\,F_{ij}\,(l^i\,n^j+m^i\,m^{*j})
\nonumber \\
\phi_2&=&F_{ij}\,m^{*i}\,n^j
\ ,
\end{eqnarray}
which contain all the information about the six components
of the electric and magnetic fields.
\par
We double expand every relevant field in $Q$ and the wave parameter $g$,
\begin{eqnarray}
&&\phi(t,r,\theta,\varphi)=\sum\limits_{n=0}^\infty\,Q^n\,\left[
\phi^{(0,n)}(r,\theta)+
g\,e^{i\,\bar\omega\,t+i\,m\,\varphi}\,\phi^{(1,n)}(r,\theta)\right]
\nonumber \\
&&\phi_i(t,r,\theta,\varphi)=\sum\limits_{n=0}^\infty\,Q^n\,\left[
\phi^{(0,n)}_i(r,\theta)
+g\,e^{i\,\bar\omega\,t+i\,m\,\varphi}\,\phi^{(1,n)}_i(r,\theta)\right]
\ ,
\ \ \ i=0,1,2
\ .
\label{(l,n)}
\end{eqnarray}
Each function of $r$ and $\theta$ at order $(1,n)$ implicitly carries
an extra integer index, $m$, and the continuous dependence on the frequency
$\bar\omega$ (both can be positive or negative).
The static metric at order zero in the ratio $Q/M$ is given by
Eq.~(\ref{metric}) with $\Delta$ replaced by $\Delta_0=r^2-2\,M\,r+\alpha^2$.
The Maxwell scalars
$\phi_0^{(0,0)}=\phi_0^{(0,1)}=\phi_2^{(0,0)}=\phi_2^{(0,1)}=\phi_1^{(0,0)}=0$
and
\begin{eqnarray}
\phi_1^{(0,1)}=-{i\over2\,(\bar\rho^*)^2}
\ .
\label{max00}
\end{eqnarray}
At order $(1,0)$ one obtains the decoupled and separable wave equations
in the Kerr background.
\par
In order to write down explicitly Eqs.~(\ref{feq}) one needs
the directional derivatives along the four null vectors (\ref{tetrad})
when acting on the wave modes displayed above.
Those can be written
\begin{eqnarray}
l^i\,\partial_i&=&{\cal D}_0
\nonumber \\
n^i\,\partial_i&=&-{\Delta\over2\,\rho^2}\,{\cal D}_0^\dagger
\nonumber \\
m^i\,\partial_i&=&{1\over\sqrt{2}\,\bar\rho}\,{\cal L}_0^\dagger
\nonumber \\
{m^*}^i\,\partial_i&=&{1\over\sqrt{2}\,\bar\rho^*}\,{\cal L}_0
\ ,
\end{eqnarray}
where
\begin{eqnarray}
{\cal D}_n&=&\partial_r+i\,{K\over\Delta}+2\,n\,{r-M\over\Delta}
\nonumber \\
{\cal D}_n^\dagger&=&\partial_r-i\,{K\over\Delta}+2\,n\,{r-M\over\Delta}
\nonumber \\
{\cal L}_n&=&\partial_\theta+\tilde Q+n\,\cot\theta
\nonumber \\
{\cal L}_n^\dagger&=&\partial_\theta-\tilde Q+n\,\cot\theta
\ ,
\label{cal D}
\end{eqnarray}
with $n$ an integer such that $n\ge 0$ and
$K\equiv(r^2+\alpha^2)\,\bar\omega+\alpha\,m$,
$\tilde Q\equiv\alpha\,\bar\omega\,\sin\theta+m\,\csc\theta$.
\subsection{Dilaton equation}
The equation for the dilaton field at order $(1,0)$ is the Klein-Gordon
equation which describes free dilaton waves on the Kerr background.
What we need is instead the wave equation at order $(1,1)$,
\begin{eqnarray}
\left[\Delta_0\,{\cal D}_1\,{\cal D}_0^\dagger
+{\cal L}_0^\dagger\,{\cal L}_1
+2\,i\,\bar\omega\,\bar\rho\right]\Phi
=-2\,\sqrt{2}\,a\,\left(\bar\rho\,\phi_1^{(0,1)}\,\Phi_1-
\bar\rho^*\,\phi_1^{(0,1)\,*}\,\Phi_1^*\right)
\ ,
\label{dila11}
\end{eqnarray}
where $\phi_1^{(0,1)}$ has been given in Eq.~(\ref{max00}),
$\Phi\equiv\phi^{(1,1)}$ and $\Phi_1\equiv\sqrt{2}\,\bar\rho^*\,\phi_1^{(1,0)}$
appearing in the current
on the RHS is one of the free Maxwell scalar waves in the Kerr background and
represents the flux of radiation emitted by the star.
It is important to notice that, due to the form of the current, ingoing
(outgoing) Maxwell waves would generate both ingoing and outgoing dilaton
waves in the same process.
\par
In the following subsections we compute $\Phi_1$ and the corresponding $\Phi$
which we shall use in section~\ref{flux} to estimate the energy absorbed
by dilaton waves from the radiation of the star.
\subsection{Maxwell waves}
As we have just shown we only need the Maxwell waves at order $(1,0)$,
thus it is convenient to introduce $\Phi_0\equiv \phi_0^{(1,0)}$ and
$\Phi_2\equiv 2\,(\bar\rho^*)^2\,\phi_2^{(1,0)}$ for which one can obtain
separate equations using Eqs.(3.1) and (3.8)
\begin{eqnarray}
&&
\left[\Delta_0\,{\cal D}_0\,{\cal D}_0^\dagger
+{\cal L}_0^\dagger\,{\cal L}_1
-2\,i\,\bar\omega\,\bar\rho\right]\Delta_0\,\Phi_0=0
\nonumber \\
&&
\left[\Delta_0\,{\cal D}_0^\dagger\,{\cal D}_0
+{\cal L}_0\,{\cal L}_1^\dagger
+2\,i\,\bar\omega\,\bar\rho\right]\Phi_2=0
\ ,
\end{eqnarray}
whose solution can be factorized as
\begin{eqnarray}
&&\Delta_0\,\Phi_0=P_{0}(r)\,S_{0}(\theta)\ ,
\ \ \ P_{0}=\Delta_0\,R_{0}
\nonumber \\
&&\Phi_2=P_{2}(r)\,S_{2}(\theta)\ ,
\ \ \ P_{2}=R_{2}
\ .
\end{eqnarray}
For any integers $1\le l$ and $|m|\le 2\,l+1$ one has
\begin{eqnarray}
S^m_{1+s,l}(\theta,\bar\omega)\,e^{i\,m\,\varphi}=
Y^m_{s=\pm 1,l}(\theta,\varphi;\bar\omega)
\ ,
\end{eqnarray}
where the $Y(\bar\omega)$ are {\em spin-weighted spheroidal harmonics\/}
\cite{X} which form a complete, orthonormal set of functions for every
(half) integer value of $s$.
They reduce to the {\em spin-weighted spherical harmonics\/},
\begin{eqnarray}
Y^m_{sl}(\theta,\varphi)=S^m_{sl}(\theta)\,e^{i\,m\,\varphi}
\ ,
\end{eqnarray}
in the limit $\bar\omega=0$ and to the usual spherical harmonics
when one has also $s=0$.
\par
Upon defining $Z\equiv\Delta^{s/2}_0\,\sqrt{r^2+\alpha^2}\,R_{1-s}$,
the radial equation can be reduced to the standard form \cite{teukolsky}
\begin{eqnarray}
Z_{,r_* r_*}+\left[{K^2-2\,i\,s\,\,(r-M)\,K
+\Delta_0\,(4\,i\,r\,s\,\bar\omega-E)\over(r^2+\alpha^2)^2}
-G^2-G_{,r_*}\right]\,Z=0
\ ,
\label{teu}
\end{eqnarray}
where $dr_*\equiv \Delta^{-1}_0\,(r^2+\alpha^2)\,dr$ is the standard
tortoise coordinate and
\begin{eqnarray}
G=s\,{r-M\over r^2+\alpha^2}+{r\,\Delta_0\over(r^2+\alpha^2)^2}
\ .
\end{eqnarray}
The quantity $E$ is the separation constant between
radial and angular equations.
This can be obtained together with similarly approximate expressions
for $S_0$ and $S_2$ upon expanding the angular equation,
\begin{eqnarray}
&&{1\over\sin\theta}\,{d\over d\theta}\left[\sin\theta\,
{dS^m_{1+s,l}\over d\theta}\right]+
\nonumber \\
&&\left[\alpha\,\bar\omega\,\cos\theta\,\left(
\alpha\,\bar\omega\,\cos\theta-2\,s\right)
-{m\over\sin^2\theta}\,\left(m+2\,s\,\cos\theta\right)
-\cot^2\theta+s-\alpha^2\,\bar\omega^2+2\,\alpha\,\bar\omega\,m
+E\right]\,S^m_{1+s,l}=0
\ ,
\end{eqnarray}
for $\alpha\,\bar\omega$ small \cite{fackerell}.
In particular, to next to leading order, one has
\begin{eqnarray}
S^m_{1+s,l}\simeq S_{1+s}^{(0)}+S_{1+s}^{(1)}\,\alpha\,\bar\omega
\ ,
\label{Sml}
\end{eqnarray}
where $S_{1+s}^{(0)}$ and $S_{1+s}^{(1)}$ are coefficients which depend
on $m$ and $l$, and
\begin{eqnarray}
E\simeq l\,(l+1)-{2\,m\,\alpha\,\bar\omega\over l\,(l+1)}
\ .
\end{eqnarray}
\par
Analytic solutions of Eq.~(\ref{teu}) for all values of $r>r_+$ are
presently available only in the form of infinite series of
hypergeometric or Coulomb functions (see \cite{mano} and Refs. therein).
It is however relatively easy to find asymptotic forms for the radial
function far away from the hole and near the horizon.
\subsubsection{Large $r$ expansion}
In the large $r$ expansion ($r\gg r_+$), the leading terms are given by
\cite{teukolsky}
\begin{eqnarray}
\begin{array}{ll}
P_{0}=A_0^{out}\,\strut\displaystyle{e^{-i\,\bar\omega\,r_*}\over r}&  \\
& {\rm (outgoing\ \ modes)} \\
P_{2}=2\,A_2^{out}\,r\,e^{-i\,\bar\omega\,r_*}&
\end{array}
\label{out_large}
\end{eqnarray}
\begin{eqnarray}
\begin{array}{ll}
P_{0}=A_0^{in}\,r\,e^{+i\,\bar\omega\,r_*}&  \\
& {\rm (ingoing\ \ modes)} \\
P_{2}=2\,A_2^{in}\,\strut\displaystyle{e^{+i\,\bar\omega\,r_*}\over r}&
\end{array}
\label{in_large}
\end{eqnarray}
where $r_*\simeq r$ and $A_{1+s}^{in/out}$ are constants whose precise
value is not important in light of the remark in the opening paragraph
of this Section.
\par
These asymptotic modes are of interest for our problem only if the
typical radius of interaction between the light and background
electromagnetic field $R\gg r_+$, that is, if one can manage to set up
a device which measures any change in luminosity with very high
accuracy.
On the other hand, present day telescopes are not able to look 
close to the horizon and the above approximation might be worth
pursuing as well.
\subsubsection{Near horizon expansion}
\label{nh}
Near the horizon the static electromagnetic field is stronger and
the current in the RHS of Eq.~(\ref{dila11}) becomes more effective.
\par
In the small $x\equiv r-r_+$ expansion physically sound boundary
conditions yield the result that only ingoing modes survive at leading order,
and the leading contributions are given by \cite{teukolsky}
\begin{eqnarray}
\begin{array}{ll}
P_{0}=A_0^{in}\,e^{i\,k\,r_*}&  \\
& {\rm (ingoing\ \ modes)} \\
P_{2}=2\,A_2^{in}\,\Delta_0\,e^{i\,k\,r_*}\ ,&
\end{array}
\label{in_close}
\end{eqnarray}
where $k=\bar\omega-m\,\alpha/2\,M\,r_+$ and $A_{s+1}^{in}$ are again
constants.
The value $\bar\omega_+\equiv \alpha/2\,M\,r_+$ is the threshold for
the onset of {\em super-radiance} and
\begin{eqnarray}
\Delta_0\simeq 2\,x\,\sqrt{M^2-\alpha^2}
\ .
\end{eqnarray}
\par
We observe that, in terms of the function $Z$ appearing in Eq.~(\ref{teu}),
the solutions displayed above are of order $x^{-s/2}$.
Since we expect the region near $r_+$ to contribute the most relevant
effects, we proceed to find an approximation to the next to leading
order in $x$, that is ${\cal O}(x^{-s/2+1})$.
In particular, one has
\begin{eqnarray}
r_*\simeq {M\,r_+\over\sqrt{M^2-\alpha^2}}\,\ln{x\over\sqrt{M^2-\alpha^2}}
\ ,
\end{eqnarray}
and, upon expanding all terms in Eq.~(\ref{teu}) around $x=0$ one obtains
\begin{eqnarray}
Z_{,r_* r_*}+\left[A+B\,e^{{\sqrt{M^2-\alpha^2}\over M\,r_+}\,r_*}\right]\,Z
=0
\ ,
\label{teu1}
\end{eqnarray}
with
\begin{eqnarray}
A&=&\left[(\bar\omega-m\,\bar\omega_+)
-i\,s\,{\sqrt{M^2-\alpha^2}\over 2\,M\,r_+}\right]^2
\nonumber \\
{\rm Re}\,B&=&{\sqrt{M^2-\alpha^2}\over M^2\,r_+}\,\left[
\alpha\,m\,(\bar\omega-m\,\bar\omega_+)
-{\sqrt{M^2-\alpha^2}\over 2\,r_+}\,\left(
E+s\,(1+s)+(1-s^2)\,{\sqrt{M^2-\alpha^2}\over M}\right)
\right]
\nonumber \\
&=&{\sqrt{M^2-\alpha^2}\over M^2\,r_+}\,\left[
\alpha\,m\,(\bar\omega-m\,\bar\omega_+)
-{\sqrt{M^2-\alpha^2}\over 2\,r_+}\,\left(
E+s+1\right)\right]
\nonumber \\
{\rm Im}\,B&=&s\,{\sqrt{M^2-\alpha^2}\over M^2\,r_+}\,\left[
(\bar\omega-m\,\bar\omega_+)\,\left(2\,\sqrt{M^2-\alpha^2}-M\right)
+\bar\omega\,\sqrt{M^2-\alpha^2}\right]
\ .
\end{eqnarray}
The change of variable
\begin{eqnarray}
z&=&2\,M\,r_+\,\sqrt{B\over M^2-\alpha^2}\,
e^{{\sqrt{M^2-\alpha^2}\over 2\,M\,r_+}\,r_*}
\nonumber \\
&=& 2\,M\,r_+\,{\sqrt{B\,x}\over (M^2-\alpha^2)^{3/4}}
\ ,
\end{eqnarray}
reduces Eq.~(\ref{teu1}) to the standard differential equation for
Bessel's functions \cite{abramo} of order
\begin{eqnarray}
\nu=s\,\left[s+i\,{2\,M\,r_+\over\sqrt{M^2-\alpha^2}}\,
(\bar\omega-m\,\bar\omega_+)\right]
\ .
\end{eqnarray}
The proper solutions are then selected by the requirement that the leading
behavior given by Eq.~(\ref{in_close}) is recovered, that is
$Z_{s=+1}\sim\Delta_0^{-1/2}$ and $Z_{s=-1}\sim\Delta_0^{1/2}$.
This yields
\begin{eqnarray}
&&Z_{s=+1}\simeq Y_{\nu}(z)={J_\nu(z)\,\cos(\nu\,\pi)-J_{-\nu}(z)
\over\sin(\nu\,\pi)}
\nonumber \\
&&Z_{s=-1}\simeq J_{\nu}(z)=\left({z\over 2}\right)\,\sum_{k=0}\,
{\left(-z^2/ 4\right)^k\over k!\,\Gamma(\nu+k+1)}
\ ,
\label{s=+1}
\end{eqnarray}
where the equalities are understood to hold only up to order $x^{-s/2+1}$
and the expression for $s=+1$ is meant to be replaced by its limit for
$\nu$ zero or integer.
\par
If we write $Z=Z^{(0)}+Z^{(1)}$ with $Z^{(0)}$ given in
Eqs.~(\ref{in_close}), up to next to leading order ($k=0,1$ in the series above)
for $s=+1$ one has
\begin{eqnarray}
Z_{s=+1}={Z_{s=+1}^{(0)}\over \sin(\nu\,\pi)}\,\left[1-
\left({z^2\over 4}\right)\,{1\over\Gamma(2-\nu)}\right]
\ ,
\label{Z1}
\end{eqnarray}
and for $s=-1$
\begin{eqnarray}
Z_{s=-1}=Z_{s=-1}^{(0)}\,\left[1-
\left({z^2\over 4}\right)\,{1\over\Gamma(\nu+2)}\right]
\ .
\label{Z-1}
\end{eqnarray}
\par
The case of $\bar\omega=m\,\bar\omega_+$ is particularly simple, since then
$\nu=1$ and
\begin{eqnarray}
&&Z_{s=+1}=-Y_1(z)
\nonumber \\
&&Z_{s=-1}=J_1(z)
\ .
\end{eqnarray}
Also $B$ simplifies to
\begin{eqnarray}
B={M^2-\alpha^2\over 2\,M^2\,r_+^2}\,\left[
1+s+E+2\,i\,s\,\bar\omega\,r_+\right]
\ .
\end{eqnarray}
\subsubsection{Completing the solution}
Once one has found $\Phi_0$ and $\Phi_2$, the solution can be completed
by computing $\Phi_1$ according to \cite{chandra}
\begin{eqnarray}
\Phi_1={1\over 2\,\bar\rho^*}\,\left[\left(
g_{+1}\,{\cal L}_1\,S_{0}-g_{-1}\,{\cal L}^\dagger_1\,S_{2}\right)
-i\,\alpha\,\left(
f_{-1}\,{\cal D}_0\,P_{2}-f_{+1}\,{\cal D}^\dagger_0\,P_{0}\right)
\right]
\ ,
\end{eqnarray}
where
\begin{eqnarray}
&&
g_{+1}={1\over C}\,\left[r\,{\cal D}_0-1\right]\,P_2
\nonumber \\
&&
g_{-1}={1\over C}\,\left[r\,{\cal D}_0^\dagger-1\right]\,P_0
\nonumber \\
&&
f_{+1}={1\over C}\,\left[\cos\theta\,{\cal L}^\dagger_1
+\sin\theta\right]\,S_2
\nonumber \\
&&
f_{-1}={1\over C}\,\left[\cos\theta\,{\cal L}_1
+\sin\theta\right]\,S_0
\ .
\end{eqnarray}
The constant $C=\sqrt{E^2-4\,\beta^2\bar\omega^2}$ and
$\beta^2=\alpha\,(\alpha+m/\bar\omega)$.
\par
For the asymptotic radial functions in Eq.~(\ref{out_large}) one
obtains
\begin{eqnarray}
\Phi_1^{out}=i\,\left[{\bar\omega\over C}\,{\cal L}_1^\dagger S_2
\,A_0^{out}
+\alpha\,\left({m\over C}\,{\cal L}_1S_0-f_{-1}\right)\,A_2^{out}
\right]\,{e^{-i\,\bar\omega\,r_*}\over r}
\ .
\label{1_out_large}
\end{eqnarray}
The coefficient multiplying $A_2^{out}$ is subleading in the small
$\alpha\,\bar\omega$ approximation, therefore in the next Section
we shall approximate $\Phi_1^{out}\sim A_2^{out}$.
The ingoing modes in the same asymptotic regime are obtained by making
use of Eq.~(\ref{in_large}),
\begin{eqnarray}
\Phi^{in}_1={i\over 2}\,\left[\alpha\,\left(
{m\over C}\,{\cal L}_1^\dagger S_2+f_{+1}\right)\,A_0^{in}
+{4\,\bar\omega\over C}\,{\cal L}_1 S_0\,A_2^{in}
\right]\,{e^{+i\,\bar\omega\,r_*}\over r}
\ .
\label{1_in_large}
\end{eqnarray}
Now it is the coefficient of $A_2^{in}$ which is subleading, thus
giving $\Phi^{in}_1\sim A_0^{in}$.
Near the horizon, from Eq.~(\ref{in_close}), one has
\begin{eqnarray}
\Phi^{in}_1={m\,\alpha^2\over 2\,\bar\rho^*\,\Delta_0}\,f_{+1}\,
A_0^{in}\,e^{i\,k\,r_*}
\ .
\label{1_in_close}
\end{eqnarray}
These are the quantities which contribute to the currents in
Eq.~(\ref{dila11}).
We note in passing that they all vanish for $\omega\to 0$ because
of the behavior of the angular functions at small $\alpha\,\bar\omega$.
\subsection{Dilaton waves}
Now we have all the ingredients to compute the dilaton waves according
to Eq.~(\ref{dila11}).
\subsubsection{Large $r$ expansion}
In the large $r$ regime, we assume
\begin{eqnarray}
\Phi^{out/in}=A_\phi^{out/in}\,S_\phi^{out/in}\,
{e^{\mp i\,\bar\omega\,r_*}\over r^{n_{out/in}}}
\ .
\end{eqnarray}
On substituting $\Phi_1^{out}$ from Eq.~(\ref{1_out_large}) into
the dilaton equation (\ref{dila11}) one obtains
\begin{eqnarray}
\left\{\begin{array}{l}
n_{out/in}=3   \\
S_\phi^{out/in}={1\over C}\,\left[(m-\cos\theta)\,{\cal L}_1
-\sin\theta\right]\,S_{0}
=s^{(1)}_{out}\,\alpha\,\bar\omega+{\cal O}(\alpha^2\,\bar\omega^2)
\\
A_\phi^{out}=-{3\over 4}\,A_\phi^{in}
=-i{\sqrt{2}\over 6}\,{a\,\alpha^2\over\alpha\,\bar\omega}\,A_2^{out}
\ ,
\end{array}\right.
\label{out/in_out_large}
\end{eqnarray}
for the outgoing and ingoing dilaton.
\par
Analogously, from $\Phi_1^{in}$ in Eq.~(\ref{1_in_large}) one finds that
the angular behavior changes while both the power of $1/r$ and the
amplitudes $A_\phi^{out/in}$ are still given by the expressions above with
$A_2^{out}$ replaced by $A_0^{in}$,
\begin{eqnarray}
\left\{\begin{array}{l}
n_{out/in}=3   \\
S_\phi^{out/in}={1\over C}\,\left[(m+\cos\theta)\,{\cal L}_1^\dagger
+\sin\theta\right]\,S_2
=s^{(1)}_{in}\,\alpha\,\bar\omega+{\cal O}(\alpha^2\,\bar\omega^2)
\\
A_\phi^{out}=-{3\over 4}\,A_\phi^{in}
=-i{\sqrt{2}\over 3}\,{a\,\alpha^2\over\alpha\,\bar\omega}\,A_0^{out}
\ .
\end{array}\right.
\label{out/in_in_large}
\end{eqnarray}
In the above, the coefficients  $s^{(1)}$ can be determined
from the expressions for $S_{1+s}$ given in Eq.~(\ref{Sml})
along with higher order terms in the small $\alpha\,\bar\omega$
approximation.
\subsubsection{Near horizon expansion}
For $r\sim r_+$ we assume
\begin{eqnarray}
\Phi^{out/in}=A_\phi^{out/in}\,S_\phi^{out/in}\,
\Delta^{n_{out/in}}\,e^{\mp i\,k\,r_*}
\ .
\end{eqnarray}
On substituting $\Phi_1^{in}$ from Eq.~(\ref{1_in_close}) into
the dilaton equation (\ref{dila11}) and expanding for small
$r-r_+$ one obtains
\begin{eqnarray}
\left\{\begin{array}{l}
n_{out/in}=0   \\
S_\phi^{out/in}={1\over C}\,\left[\cos\theta\,{\cal L}_1^\dagger
+\sin\theta\right]\,S_2
=s^{(1)}_{out/in}\,\alpha\,\bar\omega+{\cal O}(\alpha^2\,\bar\omega^2)
\\
A_\phi^{out/in}=
{\sqrt{2}\over 16}\,{a\,\alpha^2\over\alpha\,\bar\omega}\,
{A_0^{in}\over M\,r_+^3}
\ .
\end{array}\right.
\label{out/in_in_close}
\end{eqnarray}
In order to get a better estimate, one can compute $\Phi_1^{in}$
from the knowledge of the electromagnetic modes in Eqs.~(\ref{Z1})
and (\ref{Z-1}) which yields higher order terms in $r-r_+$.
\section{Energy transfer}
\label{flux}
Now that we have the amplitudes for the dilaton waves produced
by the radiation emitted by the star we can compute the energy
loss.
In order to do that we have to determine the energy carried away by
dilaton waves stimulated at a given position (local effect) and
integrate over the position of the interaction (global effect),
to wit along the paths described in section~\ref{system}.
\par
The expression for the dilaton waves obtained in the previous section are
rather involved, especially if one wants to take into account the precise
angular behavior.
For simplicity we assume that the star revolves at
$\theta=\pi/2$, as mentioned in section~\ref{system}, and then expanding
around that angle.
One can further simplify the task by focusing on the dependence of the energy
loss with respect to the frequency of the electromagnetic radiation emitted by
the star.
\par
We recall that the flux of energy carried by a given wave mode is related
to the energy-momentum tensor $T_{ij}$ by
\cite{chandra,teukolsky}
\begin{equation}
\left.{dE\over dt\,d\Omega}\right|_\infty=r^2\,T^r_{\ t}
\ ,
\label{Einfty}
\end{equation}
for $r\to+\infty$ and by \cite{chandra}
\begin{equation}
\left.{dE\over dt\,d\Omega}\right|_+
={2\,M\,r_+\,\bar\omega\over \bar\omega-\bar\omega_+}\,T_{ij}\,
\tilde l^i\,\tilde l^j
\ ,
\label{Tll}
\end{equation}
for $r\sim r_+$, where
\begin{equation}
\tilde l^i={\Delta\over 4\,M\,r_+}\,l^i
\end{equation}
and the $l^i$ are the components of the null vector
defined in Eq.~(\ref{tetrad}).
Also, for the electromagnetic field one has
\begin{eqnarray}\label{Tem}
T_{ij}&=&\left[|\phi_0|^2\,n_i\,n_j+|\phi_2|^2\,l_i\,l_j+
2\,|\phi_1|^2\,\left(l_{(i}\,n_{j)}+m_{(i}\,m^*_{j)}\right)
\right.
\nonumber \\
&&\left.
-4\,\phi_0^*\,\phi_1\,n_{(i}\,m_{j)}
-4\,\phi_1^*\,\phi_2\,l_{(i}\,m_{j)}
+2\,\phi_0^*\,\phi_2\,m_i\,m_j\right]
+{\rm c.c.}
\ ,
\end{eqnarray}
and for the dilaton
\begin{equation}
\left|T_{ij}\right|=
{Q^2\over 2}\,\left|\partial_i\Phi\,\partial_j\Phi\right|
\ ,
\end{equation}
where $Q$ is the charge generating the background Maxwell field.
\subsection{Large $r$ expansion}
\begin{figure}
\centerline{\epsfysize=220pt\epsfbox{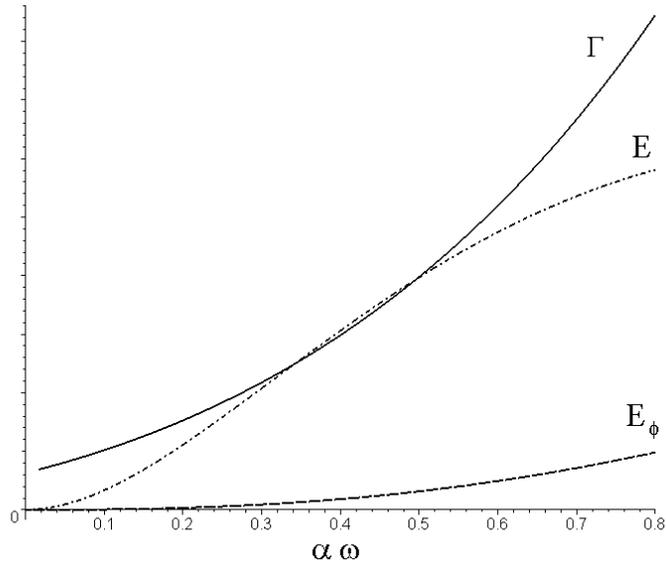}}
\caption{Qualitative behaviour of the energy flux $E_\phi$ carried by
the dilaton waves generated at large distance by electromagnetic waves
of energy $E$ for small values of $\alpha\,\bar\omega$ and $l=4$, $m=2$;
$\Gamma$ is the ratio $(E_\phi/E)$.
The vertical scale is arbitrary.}
\label{loss1}
\end{figure}

From Eqs.~(\ref{out/in_out_large}) and (\ref{out/in_in_large})
one obtains an energy flux (\ref{Einfty}) carried by the dilaton
waves generated at large distance from the horizon given by
\begin{eqnarray}
&&dE_\phi^{out/in}(out)\sim
\left(\alpha\,\bar\omega\,\,A_2^{out}\right)^2\,{Q^2\over r^4}
\nonumber \\
&&dE_\phi^{out/in}(in)\sim
\left(\alpha\,\bar\omega\,\,A_0^{in}\right)^2\,{Q^2\over r^4}
\ .
\label{f_large}
\end{eqnarray}
The (omitted) coefficients of proportionality depend on the background
parameters and $l$ and $m$ (and the dilaton coupling $a$) but
not on the frequency of the original electromagnetic wave.
To leading order in $1/r$ one also has
\begin{eqnarray}
\left.T^r_{\ t}\right|_\infty=\left\{\begin{array}{ll}
\strut\displaystyle{1\over 2\,\pi}\,{|\Phi_2|^2\over 4\,r^4}
&\ \ \ \ ({\rm outgoing\ modes})
\\
& \\
-\strut\displaystyle{1\over 8\,\pi}\,|\Phi_0|^2
&\ \ \ \ ({\rm ingoing\ modes})
\ ,
\end{array}\right.
\end{eqnarray}
from which one can compute the ratio between $dE_\phi$ and the energy
flux $dE^{out/in}$ of the original electromagnetic wave at a given
position (local effect),
\begin{eqnarray}
\Gamma(r\to\infty)\equiv {dE_\phi^{out/in}\over dE^{out/in}}
\ .
\end{eqnarray}
One finds that $\Gamma$ is in fact an increasing function of the
frequency and tends to a constant, non-zero value for zero frequency
(see Fig.~\ref{loss1}).
If $\alpha\sim M$, then for a solar mass black hole the plot
in Fig.~\ref{loss1} contains frequencies $\bar\omega<100$ KHz.
\par
For $r\sim R$ and $\Gamma$ of the order of the precision
which can be attained with existing instruments, since
$\Gamma\sim R^{-4}$, one obtains an estimate of the spatial resolution
needed to test such local effects.
It might, however, be more interesting to consider the global effect
on a given frequency mode and integrate the flux in Eq.~(\ref{f_large})
along the path followed by the wave (see Fig.~\ref{sys}),
\begin{equation}
E_\phi^{in/out}\sim\int_{r_m}^R dE_\phi^{in/out}
\sim{1\over r_m^3}-{1\over R^3}
\ .
\end{equation}
Taking for $r_m\sim 10\,M$ and $R\to\infty$ yields $\Gamma\sim 1/M^3$
with a dependence on the frequency as displayed in Fig.~\ref{loss1}.
\subsection{Near horizon expansion}
Near the horizon the energy flux (\ref{Tll}) becomes singular on the
threshold of super-radiance, therefore we shall only consider
non-super-radiant modes.
Since we also work in the small frequency approximation for the angular
part of the wave modes, the expressions we obtain in this Section hold
for
\begin{equation}
{m\,\alpha\over 2\,M\,r_+}<\bar\omega<{1\over\alpha}
\ .
\end{equation}
\par
\begin{figure}
\centerline{\epsfysize=250pt\epsfbox{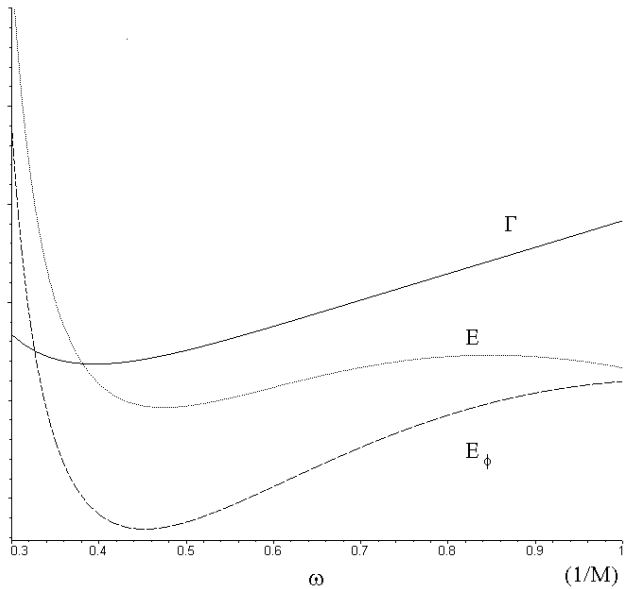}}
\caption{Qualitative behaviour of the energy flux $E_\phi$ carried
by the dilaton waves generated near the horizon by electromagnetic
waves of energy $E$ and frequency $\bar\omega$ (in units of $1/M$),
$\alpha\sim M/2$ and $l=4$, $m=2$;
$\Gamma$ is the ratio $(E_\phi/E)$.
The vertical scale is arbitrary.}
\label{loss2}
\end{figure}

For the dilaton waves in Eq.~(\ref{out/in_in_close}) one has
\begin{equation}
\left|T_{ij}\,\tilde l^i\,\tilde l^j\right|
\simeq
{1\over 2}\,\left(7\,\bar\omega^2-{11\,m\,\alpha\,\bar\omega\over
4\,M\,r_+}+{9\,m^2\,\alpha^2\over 8\,M^2\,r_+^2}\right)\,
|\Phi^{out/in}|^2
\ .
\end{equation}
The energy flux for the impinging electromagnetic wave is
\begin{equation}
\left.{dE\over dt\,d\Omega}\right|_+={\bar\omega\over8\,M\,r_+\,
(\bar\omega-\bar\omega_+)}\,{\Delta^2\over 2\,\pi}\,|\Phi_0|^2
\ .
\end{equation}
Hence the (local) ratio
\begin{eqnarray}
\Gamma(r\sim r_+)\equiv {dE_\phi^{out/in}\over dE^{in}}
\ ,
\end{eqnarray}
follows as displayed in Fig.~\ref{loss2} where it appears that $\Gamma$
begins to increase as it approaches the super-radiant threshold.
\section{Conclusion}
\label{close}
In this paper we have analyzed the energy lost by the electromagnetic
radiation emitted by a star to dilatonic waves on the background
of a companion black hole. The expansion first introduced in \cite{knd} 
has once again been used to obtain separate expressions for the Maxwell 
scalars to first order in the charge-to-mass ratio $Q/M$.
Our results apply to black holes with arbitrary angular momentum.
We have obtained expressions for the amplitudes for the dilatonic waves
generated by the incoming electromagnetic waves for large radial
distances and near the horizon.
\par
Explicit expressions for the angular part of the wave functions
were obtained in the small $\alpha\,\omega$ limit by solving the
recursion relations for the coefficients in the expansion of the
spin-weighted spherical harmonics in terms of Jacobi polynomials.
For the radial part of the wave functions we used the asymptotic
expressions given in \cite{teukolsky} at large distance.
Near the horizon we improved the relation obtained in \cite{teukolsky} 
by solving the radial wave equation to next to leading order in the
distance from the horizon.
The complete wave functions are products of the radial and angular parts
at a fixed frequency.  Thus the expressions we have obtained display 
the rate of energy loss as a function of the frequency of the incident 
electromagnetic radiation.
\par
The results we have obtained for the energy loss to the dilatonic
component of the gravitational field are qualitative because we have
considered a simplified model of the true physical system.
We assume, for example, that the observer is in the ecliptic plane
and we ignore certain relativistic effects.
However these results may prove to be a useful guide to future observers 
and should be taken into account when a detailed balance of all
the effects occurring in the vicinity of a black hole is made
and compared with the measurements.
Such an analysis is a possible way of phenomenologically supporting
or disproving the existence of a scalar (long range) component of
gravity.
\acknowledgments
This work was supported in part by the U.S. Department of Energy
under grant No.~DE-FG02-96ER40967 and by the NATO grant No.~CRG~973052.
\end{document}